\documentclass{osa-article}
\journal{oe}

\usepackage{gensymb}
\usepackage[utf8]{inputenc}
\usepackage{mathrsfs}
\usepackage{physics}
\usepackage{textcomp}
\usepackage{float}

\begin{document}

\title{A stable 2~W continuous-wave 261.5~nm laser for cooling and trapping aluminum monochloride}

\author{J. C. Shaw,\authormark{1*} S. Hannig,\authormark{2} D. J. McCarron\authormark{1$^\dagger$}}

\address{\authormark{1}Department of Physics, University of Connecticut, 196 Auditorium Road , Unit 3046, Storrs, Connecticut 06269, USA\\
\authormark{2} Agile Optic GmbH, Krähenfeld 11, 38110 Braunschweig, Germany}

\email{\authormark{*}jamie.shaw@uconn.edu} %% email address is required
\email{\authormark{$\dagger$}daniel.mccarron@uconn.edu}


\begin{abstract}
{We present a high-power tunable deep-ultraviolet (DUV) laser that uses two consecutive cavity enhanced doubling stages with LBO and CLBO crystals to produce the fourth harmonic of an amplified homebuilt external cavity diode laser. The system generates up to 2.75~W of 261.5~nm laser light with a $\sim$2~W stable steady-state output power and performs second harmonic generation in a largely unexplored high intensity regime in CLBO for continuous wave DUV light. We use this laser to perform fluorescence spectroscopy on the $X^1\Sigma\leftarrow A^1\Pi$ transition in a cold, slow beam of AlCl molecules and probe the A$^{1} \Pi\ket{v'=0,~J'=1}$ state hyperfine structure for future laser cooling and trapping experiments. This work demonstrates that the production of tunable, watt-level DUV lasers is becoming routine for a variety of wavelength-specific applications in atomic, molecular and optical physics.}
\end{abstract}

\section{Introduction}

Direct production of intense deep-ultraviolet (DUV) radiation has been historically challenging, except at specific wavelengths where rare-gas excimer lasers can be produced. While many areas in research and industry have and continue to benefit from these sources \cite{excimermicromachining,Tittel1989ExcimerLI,excimerapps}, the lack of wavelength selectivity and frequency resolution limits research in atomic, molecular, and optical physics. For example, access to single photon Rydberg excitation \cite{singlephotonrydberg}, Rydberg dressing \cite{Rydbergdressing}, tight optical lattices \cite{UVlattice1} and precise determination of physical constants \cite{Cooper:18} all require tunable, watt-level, DUV laser light with application-specific wavelengths. The development of these laser sources is also crucial to the burgeoning field of direct molecular laser cooling \cite{McCarron_2018,fitch2021laser} as several promising candidates possess strong optical transitions with high saturation intensities in the DUV, e.g., AlF, AlCl \cite{Hofs_ss_2021,HemmerlingAlCl,Centrex}, leading to large recoil velocities and the potential for significant light-mediated forces. To this end, stable, reliable and tunable lasers in the DUV with minimal complexity are highly desirable.

Although recent work shows promise for DUV semiconductor diodes \cite{DUVlaserdiode}, this technology is still in early development and will likely not realize the watt-level powers needed for many applications. A fruitful alternative approach to these direct sources is frequency conversion via nonlinear optical (NLO) crystals, which has formed the framework for many modern coherent UV and DUV lasers. In this way, we gain access to mature laser technology in the infrared (IR), spanning a broad wavelength range ($\sim$700-1100~nm) at high power ($\sim$1-100~W). Successive stages of second harmonic generation (SHG) then offer a route to coherent and tunable watt-level DUV production between $\sim$200--275~nm. Recently, others have shown this approach to be capable on $\sim$ 1~W of stable light at 243~nm and 244~nm\cite{BurkleyDUV,burkley2021stable}.

In this work, we design, construct and characterize a frequency quadrupled DUV system as a first step toward the laser cooling and trapping of aluminum monochloride (AlCl) for future studies of strongly interacting ultracold dipolar gases. Through frequency quadrupling an amplified seed laser at 1046~nm, we realize as much as $\sim$2.75~W at 261.5~nm and show a stable steady state power of $\sim$2~W over 13~hours, furthering prior demonstrations of the power scalability of this laser system \cite{BurkleyDUV,burkley2021stable,BurkleyCLBOvsBBO}. We present details for design choices alongside system performance and use this laser to perform the first spectroscopy on the AlCl A$^{1}\Pi$ state hyperfine structure using a cold, slow molecular beam.

\section{Laser System}
\begin{figure}[htbp]
\centering\includegraphics[width=13cm]{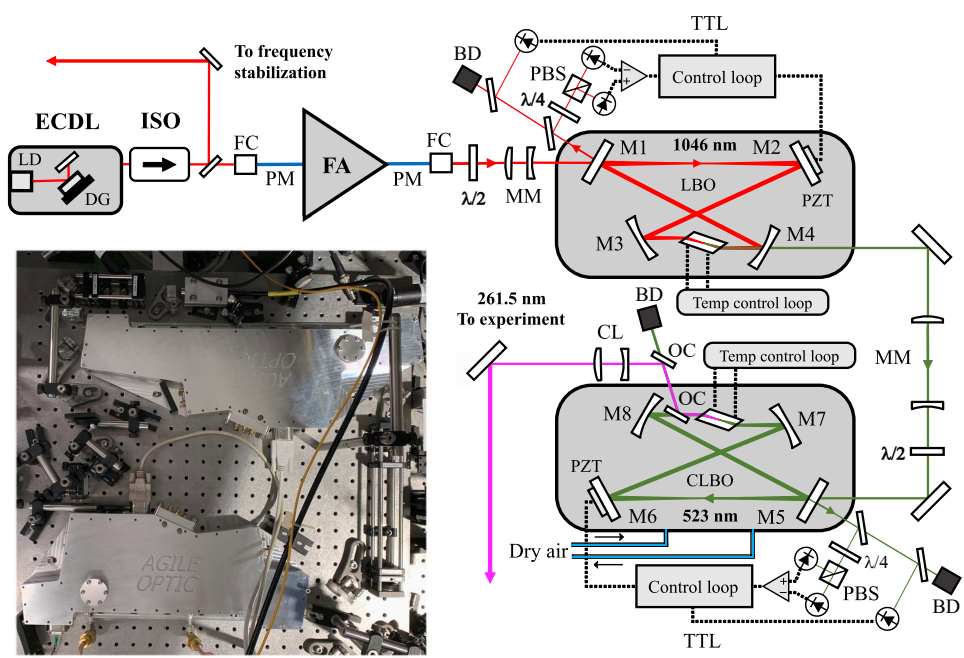}
\caption{Schematic of 261.5~nm laser system. ECDL, external cavity diode laser; LD, laser diode; DG, diffraction grating; ISO, optical isolator; FC, fiber coupler; PM, polarization maintaining fiber; FA, fiber amplifier; $\lambda/2$, half-wave plate; $\lambda/4$, quarter-wave plate; PBS, polarized beam splitter; MM, mode-matching optics;  CL, collimating lenses; OC, output coupling dichroic; BD, beam dump; PZT, piezo-electric-transducer; LBO, lithium borate; CLBO, cesium lithium borate. Inset shows a photograph of the two cavity housings without laser light present.}\label{fig1}
\end{figure}

Our system, shown in Figure \ref{fig1}, begins with a homebuilt external cavity diode laser (ECDL) emitting infrared (IR) light at 1046~nm. A portion of this light, 1-5~mW, is used to seed a fiber amplifier (IPG YAR-10-1050-LP-SF) which produces $\sim$10.75~W with a clean Gaussian mode (M$^{2}\sim$1.05). This light is sent to two cascaded resonant enhancement cavities, each containing a nonlinear crystal for second harmonic generation (SHG), producing laser light at 523 and 261.5~nm. Both cavities use a traveling wave bowtie configuration enclosed in monolithic aluminum housings, manufactured by one of the authors at Agile Optic GmbH, and are mounted on a breadboard which is vibrationally damped by Sorbothane\textsuperscript{TM}. While similar to a previous design \cite{Hannig}, we modify housing geometries to match critical parameters such as the crystal lengths and mirror radii of curvature (ROC), provided in Ref. \cite{BurkleyDUV}. For clarity, these parameters are restated in this paper.

\subsection{First doubling stage (LBO)}
The amplified IR output is coupled into the first cavity which houses a 25~mm Brewster cut Lithium Borate (LBO) nonlinear crystal used for Type I non-critical phase matching via crystal temperature. LBO is a well established NLO material at these wavelengths;  CW powers as high as 130~W at 532~nm have been demonstrated with corresponding crystal intensities of $\sim$90~MW/cm$^2$ \cite{LBO130W}, confirming suitability for high power operation. Here, a closed bowtie cavity is formed by two plano (M1 and M2) and two curved mirrors (M3 and M4). For fundamental light, M1 is a 3$\%$ transmissive input coupler (IC), while mirrors M2-M4 are high reflectors (HR) (>99.9~$\%$). Mirrors M3 and M4 have a 150~mm ROC, producing an estimated waist of 48~$\mu$m and 67~$\mu$m (the latter expanded by the Brewster incidence) at the crystal center. The waist is $\sim$2$\times$ larger than that calculated by the optimum Boyd-Kleinman focusing parameter in order to reduce thermal effects in the crystal. The generated second harmonic (SH) light at 523~nm is coupled out of the cavity via M4 (T>95~$\%$ at 523~nm). Between mirrors M1 and M2 is a secondary waist with which efficient coupling to the fundamental cavity mode is achieved via two mode-matching (MM) lenses and a half-wave plate is used to optimally set polarization angle. To maintain a precise crystal temperature, the LBO crystal is placed within a temperature-controlled metal housing with a graphite foil interface and is further enclosed by a PTFE cover to insulate it from air within the cavity housing. This element is set by an external temperature controller to $\sim$168$^{\circ}$C. When optimized, at 10.75~W input power,  we realize a $\sim$95~$\%$ coupling into the cavity, measured by reflected contrast, and a circulating power of $\sim$220~W, inferred from IR light leaked through M3. While not shown in Figure \ref{fig1}, the Brewster cut of the crystal leads to $\ll$1~$\%$ and $\sim$18$\%$ reflections of the fundamental and SH light, respectively, off the first and second interfaces, which are used to monitor both circulating power and SH conversion. The out-coupled light from this cavity is then mode-matched to the second cavity's in-coupling waist using two spherical lenses, with polarization rotated again by a half-wave plate.

\subsection{Second doubling stage (CLBO)}

\subsubsection{Crystal selection }
Efficient frequency conversion of the high power visible light from the first cavity into the DUV places strict requirements on the crystal of the second cavity. While the LBO crystal used in the first cavity is transmissive over a range extending well into the DUV, its low effective nonlinear coefficient and inadequate phase matching below $\sim$700~nm make it a poor visible light doubler. As of now, Barium Borate (BBO) and Cesium Lithium Borate (CLBO) are generally considered the best options for SHG to DUV wavelengths. At the fundamental wavelength for the second cavity, 523~nm, CLBO has a $\sim$1.5$\times$ smaller nonlinear coefficient than BBO but doubles more efficiently due to its $\sim$2.5$\times$ smaller walk-off angle \cite{BurkleyCLBOvsBBO, CLBOvsBBOdeff}. 

When placed in an enhancement cavity generating high circulating powers and large intensities within the crystal, doubling efficiency is prone to degradation. For BBO, this degradation stems from both high linear and nonlinear absorption \cite{BBOlinearabs,BBOvsreprate}. In general, this leads to self heating of the crystal along with a corresponding variation in phase matching along the beam's path which is further exacerbated by the formation of absorption sites from UV light. This in turn sets a limit on crystal intensities and attainable DUV output powers to well below those causing permanent laser induced damage. While this effect is less dramatic in CLBO due to $\sim$10$\times$ lower linear and nonlinear absorption and a $\sim$2$\times$ higher temperature bandwidth \cite{CLBOtwophotonabs,CLBOvsBBOdegradation,CLBOvsBBOpulsed}, UV-induced degradation has been shown to occur due to a separate mechanism of photorefraction \cite{CLBOUVinduceddegandphotorefractdamage}. However, by comparison, UV-induced degradation in CLBO should become significant for average intensities which are $\sim5\times$ higher than BBO, at $\sim$10~kW/cm$^2$, while still more than 20$\times$ lower than CLBO's absorption based permanent laser induced damage threshold (LIDT) \cite{CLBOvsBBOdegradation,CLBOwateranddry}. This highlights CLBO as being the better crystal option for efficient SHG over the full range of desired output powers from the second cavity.

\subsubsection{Cavity parameters}

For the second cavity, we use a 10~mm Brewster cut CLBO crystal which is Type I critically phase-matched due to it’s weak temperature dependent index.  Since CLBO is strongly hygroscopic, a similar housing to the first cavity is used to maintain a $\sim$150$^\circ$C crystal temperature which prevents water diffusion into the crystal lattice. This is crucial to increasing the threshold for UV-induced degradation and has similarly been shown to increase the LIDT 3-fold \cite{CLBOwater}. At these elevated temperatures, conversion efficiencies are also known to improve due an alleviation of distortions to the refractive index caused by both crystal processing and thermally induced shock associated with laser absorption \cite{CLBObible,CLBOannealing}. In addition, a positive pressure environment is created within the cavity housing with a continuous 10~sccm flow of dry air to provide a clean environment for the expected high circulating powers as well as to further protect CLBO from water diffusion\cite{CLBOwateranddry}.

Following Refs. \cite{BurkleyDUV} and \cite{5WUV}, two mirrors with a 200~mm ROC form a waist in the crystal center of 47~$\mu$m and 76~$\mu$m (the latter expanded by the Brewster incidence). This waist is $\sim$3$\times$ larger than the theoretical optimum to reduce thermal stress and the resulting distortion in the crystal, similar to the first cavity. The mirrors consist of a 2.5~$\%$ IC and 3 HR mirrors of >99.95~$\%$ for the fundamental light. Despite a slight ellipticity in the beam exiting the first cavity, we find that two spherical lenses are sufficient to provide high in-coupling with a measured $\sim$90~$\%$ reflected contrast. This leads to a measured intracavity circulating power of $\sim$140~W at 523~nm. The SH light at 261.5~nm is coupled out of the cavity using a HR for DUV light which is AR coated on its first surface and set to Brewster's angle relative to its second for the fundamental wavelength (OC in Figure \ref{fig1}). The small fraction of visible light which is coupled out of the cavity along with the DUV light is removed using a second Brewster's angled HR; as used within the cavity. Following the second HR, the DUV light is measured to be >99~$\%$ pure. Like the first cavity, circulating power and SHG are monitored via the two reflecting facets of the crystal.

\subsection{Electronics}
Stabilizing the cavity lengths to the amplified ECDL seed light is performed using a H{\"a}nsch-Couillaud (HC) lock \cite{HClock}. The simplicity and low cost of our homebuilt ECDL design comes with an inherent sensitivity to current noise, serving as the dominate frequency noise source over short timescales. The result of this can be seen as fast amplitude fluctuations on the SH light of each cavity due to insufficient feedback bandwidth needed to follow this frequency jitter. To address this short-term noise we use a low noise current source (Koheron DRV110), modified for feedforward to enable improved mode-hop-free tuning ranges in the DUV greater than 100~GHz.  Feedback bandwidths of $\sim$30~kHz (inferred through both in-loop and out-of-loop measurements) are achieved with cavity-specific loop-filters designed to address nearby resonances in the piezo-electric-transducers (PZTs) of the two cavities. While this method is effective, future designs will incorporate techniques discussed in \cite{YostPZT} to suppress PZT resonances and ease demand on feedback loop design.

Both cavities employ the HC lock because the error signal line-shape is well-suited for enhancement cavities. Unlike a cavity stabilized laser, the longitudinal cavity mode of an enhancement cavity is a free parameter, providing an inherent automatic relocking mechanism for error signals with a single sign above and below resonance (e.g., those derived from HC or transmission-based dither locks). This results in a single feedback direction between modes and offers cavity locks which are resilient to external perturbations without the need to ramp the cavity length to reacquire lock. We leverage this auto-relock property by supplying each control loop with a simple analog circuit composed of a window comparator to toggle the feedback loop when the PZT voltage runs out of the range set by a predefined threshold. This enables the cavities to maintain lock indefinitely (with brief $\lesssim$100~ms relocks) using a single PZT (max displacement of $\sim$9~$\mu$m). To fully automate cavity locking, the state of each feedback loop is controlled by a master switch driven by the presence of fundamental laser light reflected from each cavity (see Figure \ref{fig1}), therefore reducing required user input to tuning the ECLD and turning on (off) the fiber amplifier at the start (end) of use.

\section{Performance}

\begin{figure}[h]
\centering\includegraphics[width=13cm]{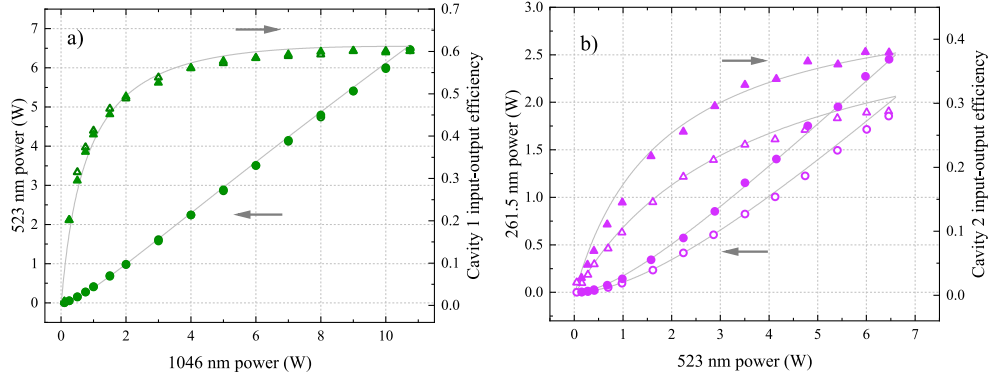}
\caption{Measured SHG output power and conversion efficiency for the first (a) and second (b) cavity. Initial powers and conversion efficiencies are indicated by filled circles and triangles, respectively, while open circles and triangles show power and conversion efficiency for steady state operation. Simulated power curves and corresponding efficiencies are provided for each data set using measured and known cavity parameters \cite{Burkelythesis}.}
\label{fig2}
\end{figure}

 The input-output powers of the first doubling cavity are shown in Figure \ref{fig2}a shortly after turn on and then again following two hours of steady state behavior. No power degradation is observed during operation. With 10.75~W of IR power, the cavity produces 6.5~W of SH light ($\gtrsim$60~$\%$ conversion efficiency) from the cavity output. Accounting for Fresnel reflection of the SH light off the crystal and that off the output coupler (M4), we infer a high, $\sim$72~$\%$, total conversion efficiency from the LBO crystal. The performance of this cavity is robust, although we measure the cavity output to increase slowly ($\sim$5~minutes) to full power from turn-on. This warm-up appears to be independent of cavity alignment and is repeated following a momentary turn-off of the system, suggesting a thermal effect within the LBO crystal. This effect is not seen in Figure \ref{fig2}a as each data point was taken after warm-up had occurred. In tandem with this warm-up in power, the cavity optical path length is seen to increase monotonically until a steady state is reached. This is identified by a corresponding increase in PZT length required to compensate for an increasing effective optical path length of the cavity. 

Example input-output power curves for the second cavity are shown in Figure \ref{fig2}b. Similar to the first cavity, the second cavity shows a monotonic change in power during warm up, as well as an associated increase in cavity length. However, in contrast to the first cavity, the produced DUV light typically reduces by $\sim$20~$\%$ as steady state is approached during the first $\sim$15~minutes of operation. The initial DUV powers and those in steady state are separately shown in Figure \ref{fig2}b. While we observe a slight indication of self heating within the crystal, evidenced by asymmetric transmission peak widths when scanning the cavity, as reported using BBO \cite{253nmBBO,229nmBBOCdMOT}, we see no clear threshold power beyond which conversion efficiency degrades, suggesting thermal dephasing is minimal. Similarly, no evidence is found for damage to the out-coupling dichroic (OC) and thus we attribute any observed UV degradation to changes in the crystal. To isolate this effect from permanent crystal damage we turn off the pump laser and allow the crystal to cool for $\sim$10~minutes. Without realignment, we find this amount of time is generally sufficient to recover $\ge$90~$\%$ of the initial power, after which power decay to the previous minimum value is seen in a similar time. The remaining power is consistently recovered following longer periods without operation (hour to day timescales). This observed long-term recovery is consistent with a process of reannealing known to take place in CLBO when held at temperatures $\gtrsim$120$^{\circ}$C over extended periods of time \cite{CLBOannealing}. Through this process, we have seen performance recover even beyond its initial power, following one week without operating, while holding both crystals at their optimal temperatures (T$_{\rm{LBO}}=168^{\circ}$C, T$_{\rm{CLBO}}=150^{\circ}$C) and no system modifications. With the IR pump at a maximum power, and the first cavity producing $\sim$7~W at 523~nm, we have measured outputs as high as $\sim$2.75~W in the DUV with a steady state power of $\sim$2~W. Accounting for Fresnel reflection, this corresponds to a doubling efficiency from the second cavity as high as $\sim$ 50~$\%$, settling to $\sim$ 30~$\%$ in the steady state.

The DUV powers generated in this work correspond to an estimated SH intensity as high as $\sim$90~kW/cm$^2$ within the crystal center, nearly 10$\times$ above the time-average threshold for UV-induced degradation suggested for CLBO (and 2$\times$ higher than was previously studied in \cite{CLBOvsBBOdegradation}). This suggests the observed drop in output power is largely a result of photorefractive UV-induced degradation known to be present in CLBO. While we note that this regime of DUV average intensities is largely unexplored with CW light, we suspect that thermal effects may also contribute. Resistance to both mechanisms of degradation is known to be increased with improved crystal quality. In contrast with the behavior of our system, the work reported in \cite{BurkleyDUV} showed no degradation at $\sim$1~W, this could be due to inherent variations in crystal quality. Nonetheless, higher purity crystals as well as those with appropriate dopings are both active areas of research \cite{CLBOphotorefractivedamageAl,CLBOaldoping2017}, offering a possible means for alleviating this degradation in future systems.

\begin{figure}[b!]
\includegraphics[width=13cm]{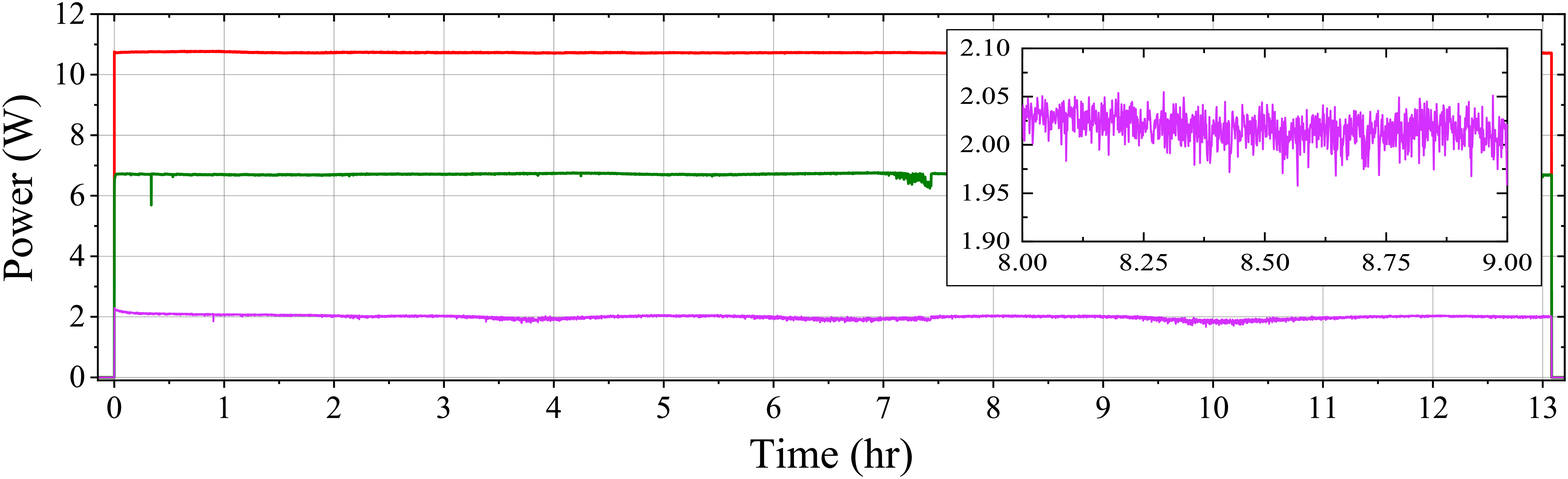}

\caption{Typical power stability observed during an operational day. Red, green and purple show measured output powers of 1046~nm, 523~nm and 261.5~nm light, respectively. The inset shows power fluctuations over hour timescales.}
\label{fig3}
\end{figure}

Following the initial $\sim$15~minutes required to reach steady state, a constant DUV power is observed without any further reduction over long timescales.  Figure \ref{fig3} shows continuous laser performance over a standard operating day from initial turn-on. Stable output power from the commercial fiber amplifier in the IR is transferred into the DUV (stable to <5~$\%$ over hour timescales). Over the range of cavity lengths available from the PZT, a change in cavity alignment, and resulting change in SH power, is observed from each cavity as the PZT tracks length changes throughout the day. We minimize this effect by limiting the allowed length tuning of each cavity to $\sim$ $\lambda$ and $\sim$ $\lambda/2$ (V$_{\rm{PZT}}\sim$15~V) using the window comparators discussed previously. Auto-relocking can be seen in Figure \ref{fig3} as repeated small jumps in power during the first hour of operation as the cavities lock to successively higher longitudinal modes. The fluctuations seen over longer timescales are periodic and correlated with lab temperature cycles. This stems from a temperature dependent set-point in our temperature controllers, particularly impactful to the doubling efficiency of the first cavity, but propagating to the second, where it is seen more clearly. This can be easily improved, however, at the time of writing, has not yet been addressed. Day to day, the output power of this system is robust, with minor optimization in alignment made to both cavities every other week. Over four months of semi-daily operation at full power ($\sim$500~hours) we find no indications of irrecoverable loss to fourth harmonic generation from this system and frequently produce initial powers of $\sim$ 2.5~W and steady state powers of $\sim$2~W.

\section{Demonstration using AlCl}

 To demonstrate the frequency tuning ability and coherence of this DUV laser, we probe a cold and slow molecular beam of AlCl produced via laser ablation from a cryogenic buffer gas beam source, previously described in \cite{Continuousbeam}. To provide frequency stabilization for this measurement, a portion of the ECDL light is simultaneously sent to a commercial wavemeter, acting as a coarse ($\sim$1~GHz) frequency ruler, and a scanning Fabry-Perot cavity referenced to a homebuilt polarization stabilized Helium-Neon (HeNe) laser, of similar design to that described in \cite{HallHeNe}. This provides frequency resolution and stability below 1 MHz over a frequency range limited by our diode laser. Lacking an identical DUV laser for comparison, we place a $\sim$1~MHz upper bound on laser linewidth in the visible using the known Free Spectral Range (FSR) and measured finesse of the second cavity. With a tight cavity lock, this corresponds to a linewidth estimate in the DUV of $\sim$2~MHz.

\begin{figure}[H]
\centering\includegraphics[width=10cm]{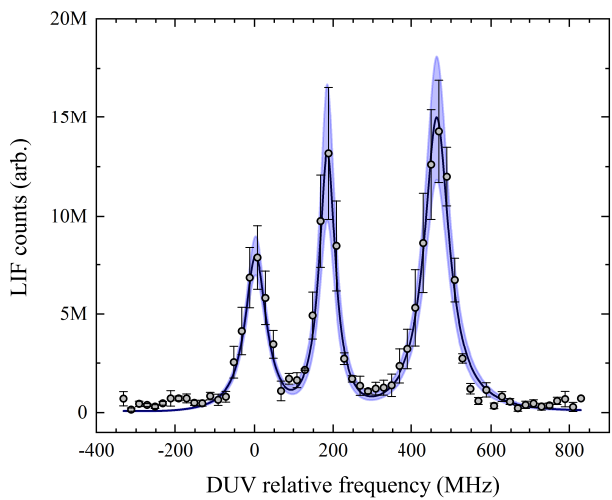}
\caption{Measured fluorescence spectrum for the $X^1\Sigma\ket{v''=0,~J''=0}\leftarrow A^1\Pi\ket{v'=0,~J'=1}$ transition of AlCl. The data (circles) shows a partly resolved hyperfine structure and a fit (solid line) is used to make a comparison of the measurement resolution to the predicted natural linewidth. The shaded region indicates the 1$\sigma$ confidence interval.}
\label{fig4}
\end{figure}

After isolating the DUV light from leaked visible light, and following two collimating lenses, a small fraction is sent transverse to the molecular beam to excite molecules with a low intensity of 0.1~mW/cm$^2$ to limit power broadening. Transverse Doppler broadening is similarly restricted to <10~MHz using an aperture on the molecular beam. Following excitation, approximately $\simeq$2~$\%$ of the resulting laser-induced fluorescence (LIF) from the molecules is collected and imaged onto an electron multiplying charge-coupled device (EMCCD) camera. Using the ECDL, we tune our laser to emit at 261.5~nm and scan over the $X^1\Sigma\ket{v''=0,~J''=0}\leftarrow A^1\Pi\ket{v'=0,~J'=1}$ transition, where $v$ and $J$ are vibrational and rotational quantum numbers, respectively, and $''$ ($'$) denotes the ground (excited) state. While the excited state hyperfine structure is unknown, previous work \cite{AlClhfs}, and calculations to appear in a future publication, indicate that the ground state hyperfine structure is unresolved, owing to the large predicted natural linewidth of $\Gamma =$ 2$\pi\times$(32~MHz) for this transition. As a result, a frequency scan across this transition provides a direct measurement of the A$^{1} \Pi\ket{v'=0,~J'=1}$ state hyperfine structure for the first time. As shown in Figure \ref{fig4}, the hyperfine structure, which consists of twelve hyperfine states, is only partly resolved. Assuming negligible Doppler broadening, a fit of three Lorentzians reveals that the narrowest feature at $\simeq$180~MHz has a linewidth of $\simeq$50~MHz. Here, unresolved hyperfine structure limits resolution rather than the predicted natural linewidth of the A$^1\Pi$ state.

\section{Outlook}
We have designed, constructed and characterized a tunable DUV laser system consisting of an amplified homebuilt-ECDL and two consecutive cavity enhanced frequency doubling stages using LBO and CLBO crystals. The system is shown to generate up to 2.75~W of 261.5~nm laser light with a $\sim$2~W stable steady state output power and a corresponding $\sim$19~$\%$ total conversion efficiency. An analog auto-locking circuit, applied to both cavities, limits user input to setting a fiber amplifier and ECDL tuning, enabling a maintained lock over the large $>$100~GHz mode-hop-free tuning range provided by the system. We have used this laser to perform the first spectroscopy on the $A^{1}\Pi$ hyperfine structure of AlCl and find that in the $A^1\Pi\ket{v'=0,~J'=1}$ state this structure is only partly resolved. This spectroscopy will provide key information needed for future work laser cooling and trapping this molecule.

The power levels generated in this work, to the best our knowledge, represent a record high for CW laser light below 266~nm, demonstrating the proposed power scalability for this system \cite{BurkleyDUV}. At these corresponding crystal intensities, we observe a recoverable UV-induced degradation in output power, consistent with observations reported with pulsed laser systems using CLBO \cite{CLBOvsBBOdegradation, CLBOdegradation2,CLBOUVinduceddegandphotorefractdamage}. While higher DUV powers should be possible (using higher IR pump powers, and/or a longer $\sim$15~mm CLBO crystal; shown for 266~nm in \cite{15mmCLBO}), addressing UV-induced degradation and thermal dephasing likely requires higher crystal quality than that used in this work. Marked progress over the last 20 years with CLBO, however, indicates this is an intense field of research motivated by industrial applications. We therefore anticipate continued crystal developments and resulting improved performance for future systems. This work builds on significant recent progress in realizing high power DUV laser sources \cite{5WUV,BurkleyDUV,229nmBBOCdMOT,253nmBBO} and highlights that the production of robust watt-level CW DUV lasers is becoming routine, enabling access to a myriad of applications in atomic, molecular and optical physics.

\begin{backmatter}
\bmsection{Funding}
We acknowledge financial support from the NSF (CAREER Award No. 1848435) and the University of Connecticut, including a Research Excellence Award from the Office of the Vice President for Research.

\bmsection{Acknowledgments}
We thank D. DeMille, D. Hanneke and L. Hunter for helpful input early on in this work and are grateful to D. Yost for insightful discussions and suggestions during the design and construction phases. We also thank P. Gould for his comments on the manuscript.

% \bmsection{Disclosures}
% The authors declare no conflicts of interest.

\bmsection{Data availability} Data underlying the results presented in this paper are not publicly available at this time but may be obtained from the authors upon reasonable request.

\end{backmatter}

% \section{Conclusion}
% After proofreading the manuscript, compress your .tex manuscript file and all figures (which should be in EPS or PDF format) in a ZIP, TAR or TAR-GZIP package. All files must be referenced at the root level (e.g., file \texttt{figure-1.eps}, not \texttt{/myfigs/figure-1.eps}). If there are supplementary materials, the associated files should not be included in your manuscript archive but be uploaded separately through the Prism interface.

%%%%%%%%%%%%%%%%%%%%%%% References %%%%%%%%%%%%%%%%%%%%%%%%%

% Add references with BibTeX or manually.
% \cite{Zhang:14,OSA,FORSTER2007,Dean2006,testthesis,Yelin:03,Masajada:13,codeexample}

%%%%%%%%%% If using BibTeX:
\bibliography{sample}

\begin{thebibliography}{10}
\newcommand{\enquote}[1]{``#1''}

\bibitem{excimermicromachining}
H.~K. Tönshoff, D.~Hesse, and J.~Mommsen, \enquote{Micromachining using
  excimer lasers,} {\protect\JournalTitle{CIRP Annals}} \textbf{42}, 247--251
  (1993).

\bibitem{Tittel1989ExcimerLI}
F.~Tittel, I.~Saidi, G.~Pettit, P.~Wisoff, and R.~Sauerbrey, \enquote{Excimer
  lasers in medicine,} in \emph{Photonics West - Lasers and Applications in
  Science and Engineering,}  (1989).

\bibitem{excimerapps}
S.~Beggs, J.~Short, M.~Rengifo-Pardo, and A.~Ehrlich, \enquote{Applications of
  the excimer laser: A review.} {\protect\JournalTitle{Dermatol Surg.}}
  \textbf{41}, 1201--1211 (2015).

\bibitem{singlephotonrydberg}
J.~Wang, J.~Bai, J.~He, and J.~Wang, \enquote{Single-photon cesium {R}ydberg
  excitation spectroscopy using 318.6~nm {UV} laser and room-temperature vapor
  cell,} {\protect\JournalTitle{Opt. Express}} \textbf{25}, 22510--22518
  (2017).

\bibitem{Rydbergdressing}
G.~Pupillo, A.~Micheli, M.~Boninsegni, I.~Lesanovsky, and P.~Zoller,
  \enquote{Strongly correlated gases of {R}ydberg-dressed atoms: Quantum and
  classical dynamics,} {\protect\JournalTitle{Phys. Rev. Lett.}} \textbf{104},
  223002 (2010).

\bibitem{UVlattice1}
S.~Mejri, J.~J. McFerran, L.~Yi, Y.~L. Coq, and S.~Bize, \enquote{Ultraviolet
  laser spectroscopy of neutral mercury in a one-dimensional optical lattice,}
  {\protect\JournalTitle{Phys. Rev. A}} \textbf{84}, 032507 (2011).

\bibitem{Cooper:18}
S.~F. Cooper, Z.~Burkley, A.~D. Brandt, C.~Rasor, and D.~C. Yost,
  \enquote{Cavity-enhanced deep ultraviolet laser for two-photon cooling of
  atomic hydrogen,} {\protect\JournalTitle{Opt. Lett.}} \textbf{43}, 1375--1378
  (2018).

\bibitem{McCarron_2018}
D.~McCarron, \enquote{Laser cooling and trapping molecules,}
  {\protect\JournalTitle{Journal of Physics B: Atomic, Molecular and Optical
  Physics}} \textbf{51}, 212001 (2018).

\bibitem{fitch2021laser}
N.~J. Fitch and M.~R. Tarbutt, \enquote{Laser cooled molecules,}  (2021).

\bibitem{Hofs_ss_2021}
S.~Hofsäss, M.~Doppelbauer, S.~C. Wright, S.~Kray, B.~G. Sartakov,
  J.~P{\'{e}}rez-R{\'{\i}}os, G.~Meijer, and S.~Truppe, \enquote{Optical
  cycling of {A}l{F} molecules,} {\protect\JournalTitle{New Journal of
  Physics}} \textbf{23}, 075001 (2021).

\bibitem{HemmerlingAlCl}
J.~R. Daniel, C.~Wang, K.~Rodriguez, B.~Hemmerling, T.~N. Lewis, C.~Bardeen,
  A.~Teplukhin, and B.~K. Kendrick, \enquote{Spectroscopy on the
  ${A}^{1}\mathrm{\ensuremath{\Pi}}\ensuremath{\leftarrow}{X}^{1}\mathrm{\ensuremath{\Sigma}}^{+}$
  transition of buffer-gas-cooled {A}l{C}l,} {\protect\JournalTitle{Phys. Rev.
  A}} \textbf{104}, 012801 (2021).

\bibitem{Centrex}
J.~O. Grasdijk, O.~Timgren, J.~Kastelic, T.~Wright, S.~K. Lamoreaux, D.~P.
  DeMille, K.~Wenz, M.~Aitken, T.~Zelevinsky, T.~Winick, and D.~Kawall,
  \enquote{Centrex: A new search for time-reversal symmetry violation in the
  $^{205}${T}l nucleus,} {\protect\JournalTitle{Quantum Science and
  Technology}}  (2021).

\bibitem{DUVlaserdiode}
Z.~Zhang, M.~Kushimoto, T.~Sakai, N.~Sugiyama, L.~J. Schowalter, C.~Sasaoka,
  and H.~Amano, \enquote{A 271.8 nm deep-ultraviolet laser diode for room
  temperature operation,} {\protect\JournalTitle{Applied Physics Express}}
  \textbf{12}, 124003 (2019).

\bibitem{BurkleyDUV}
Z.~Burkley, A.~D. Brandt, C.~Rasor, S.~F. Cooper, and D.~C. Yost,
  \enquote{Highly coherent, watt-level deep-{UV} radiation via a
  frequency-quadrupled {Y}b-fiber laser system,} {\protect\JournalTitle{Appl.
  Opt.}} \textbf{58}, 1657--1661 (2019).

\bibitem{burkley2021stable}
Z.~Burkley, L.~de~Sousa~Borges, B.~Ohayon, A.~Golovozin, J.~Zhang, and
  P.~Crivelli, \enquote{Stable high power deep-{UV} enhancement cavity in
  ultra-high vacuum with fluoride coatings,} {\protect\JournalTitle{Opt.
  Express}} \textbf{29}, 27450--27459 (2021).

\bibitem{BurkleyCLBOvsBBO}
Z.~Burkley, C.~Rasor, S.~F. Cooper, A.~D. Brandt, and D.~C. Yost, \enquote{Yb
  fiber amplifier at 972.5 nm with frequency quadrupling to 243.1 nm,}
  {\protect\JournalTitle{Applied Physics B}} \textbf{123} (2016).

\bibitem{Hannig}
S.~Hannig, J.~Mielke, J.~A. Fenske, M.~Misera, C.~O. N.~Beev, and P.~O.
  Schmidt, \enquote{A highly stable monolithic enhancement cavity for second
  harmonic generation in the ultraviolet,} {\protect\JournalTitle{Review of
  Scientific Instruments}} \textbf{89}, 13106 (2018).

\bibitem{LBO130W}
T.~Meier, B.~Willke, and K.~Danzmann, \enquote{Continuous-wave single-frequency
  532 nm laser source emitting 130 {W} into the fundamental transversal mode,}
  {\protect\JournalTitle{Opt. Lett.}} \textbf{35}, 3742--3744 (2010).

\bibitem{CLBOvsBBOdeff}
W.~L. Zhou, Y.~Mori, T.~Sasaki, and S.~Nakai, \enquote{High-efficiency
  intracavity continuous-wave ultraviolet generation using
  {C}s{L}i{B}$_{6}${O}$_{10}$, $\beta$-{B}a{B}$_2${O}$_4$, and
  {L}i{B}$_3${O}$_5$,} {\protect\JournalTitle{Optics Communication}}
  \textbf{123}, 583--586 (1996).

\bibitem{BBOlinearabs}
M.~Watanabe, K.~Hayasaka, H.~Imajo, J.~Umezu, and S.~Urabe, \enquote{Generation
  of continuous-wave coherent radiation tunable down to 190.8~nm in
  $\beta$-{B}a{B}$_2${O}$_4$,} {\protect\JournalTitle{Applied Physics B}}
  \textbf{53}, 11--13 (1991).

\bibitem{BBOvsreprate}
M.~Takahashi, A.~Osada, A.~Dergachev, P.~F. Moulton, M.~Cadatal-Raduban,
  T.~Shimizu, and N.~Sarukura, \enquote{Effects of pulse rate and temperature
  on nonlinear absorption of pulsed 262~nm laser light in
  $\beta$-{B}a{B}$_2${O}$_4$,} {\protect\JournalTitle{Japanese Journal of
  Applied Physics}} \textbf{49}, 080211 (2010).

\bibitem{CLBOtwophotonabs}
T.~Kamimura, R.~Nakamura, H.~Horibe, M.~Nishioka, M.~Yamamoto, M.~Yoshimura,
  Y.~Mori, T.~Sasaki, and K.~Yoshida, \enquote{Characterization of two-photon
  absorption related to the enhanced bulk damage resistance in
  {C}s{L}i{B}$_6${O}$_{10}$ crystal,} {\protect\JournalTitle{Japanese Journal
  of Applied Physics}} \textbf{44}, L665--L667 (2005).

\bibitem{CLBOvsBBOdegradation}
K.~Takachiho, M.~Yoshimura, Y.~Takahashi, M.~Imade, T.~Sasaki, and Y.~Mori,
  \enquote{Ultraviolet laser-induced degradation of {C}s{L}i{B}$_6${O}$_{10}$,
  $\beta$-{B}a{B}$_2${O}$_4$,} {\protect\JournalTitle{Optical Materials
  Express}} \textbf{4}, 559--567 (2014).

\bibitem{CLBOvsBBOpulsed}
D.~J. Brown and M.~J. Withford, \enquote{High-average-power (15 {W}) 255~nm
  source based on second-harmonic generation of a copper laser master
  oscillator power amplifier system in {C}esium {L}ithium {B}orate,}
  {\protect\JournalTitle{Opt. Lett.}} \textbf{26}, 1885--1887 (2001).

\bibitem{CLBOUVinduceddegandphotorefractdamage}
M.~Yoshimura, Y.~Oeki, Y.~Takahashi, H.~Adachi, and Y.~Mori,
  \enquote{Ultraviolet laser-induced degradation of {C}s{L}i{B}$_6${O}$_{10}$,}
  in \emph{Advanced Solid State Lasers,}  (Optical Society of America, 2015),
  p. AM5A.11.

\bibitem{CLBOwateranddry}
T.~Kawamura, M.~Yoshimura, Y.~Honda, M.~Nishioka, Y.~Shimizu, Y.~Kitaoka,
  Y.~Mori, and T.~Sasaki, \enquote{Effect of water impurity in
  {C}s{L}i{B}$_6${O}$_{10}$ crystals on bulk laser-induced damage threshold and
  transmittance in the ultraviolet region,} {\protect\JournalTitle{Appl. Opt.}}
  \textbf{48}, 1658--1662 (2009).

\bibitem{CLBOwater}
M.~Nishioka, A.~Kanoh, M.~Yoshimura, Y.~Mori, T.~Sasaki, T.~Katsura, T.~Kojima,
  and J.~Nishimae, \enquote{Improvement in {UV} optical properties of
  {C}s{L}i{B}$_6${O}$_{10}$ by reducing water molecules in the crystal,}
  {\protect\JournalTitle{Japanese Journal of Applied Physics}} \textbf{44},
  L699--L700 (2005).

\bibitem{CLBObible}
T.~Sasaki, Y.~Mori, M.~Yoshimura, Y.~K. Yap, and T.~Kamimura, \enquote{Recent
  development of nonlinear optical borate crystals: key materials for
  generation of visible and {UV} light,} {\protect\JournalTitle{Materials
  Science and Engineering: R: Reports}} \textbf{30}, 1--54 (2000).

\bibitem{CLBOannealing}
Y.~K. Yap, T.~Inoue, H.~Sakai, Y.~Kagebayashi, Y.~Mori, T.~Sasaki, K.~Deki, and
  M.~Horiguchi, \enquote{Long-term operation of {C}s{L}i{B}$_6${O}$_{10}$ at
  elevated crystal temperature,} {\protect\JournalTitle{Opt. Lett.}}
  \textbf{23}, 34--36 (1998).

\bibitem{5WUV}
J.~Sakuma, Y.~Asakawa, and M.~Obara, \enquote{Generation of 5 {W} deep-{UV}
  continuous-wave radiation at 266 nm by an external cavity with a
  {C}s{L}i{B}$_6${O}$_{10}$ crystal,} {\protect\JournalTitle{Opt. Lett.}}
  \textbf{29}, 92--94 (2004).

\bibitem{HClock}
T.~W. H{\"a}nsch and B.~Couillaud, \enquote{Laser frequency stabilization by
  polarization spectroscopy of a reflecting reference cavity,}
  {\protect\JournalTitle{Opt. Comm.}} \textbf{35}, 441--444 (1980).

\bibitem{YostPZT}
T.~C. Briles, D.~C. Yost, A.~Cingöz, J.~Ye, and T.~R. Schibli, \enquote{Simple
  piezoelectric-actuated mirror with 180 k{H}z servo bandwidth,}
  {\protect\JournalTitle{Opt. Express}} \textbf{18}, 9739--9746 (2010).

\bibitem{Burkelythesis}
Z.~N. Burkley, \enquote{High-power deep-{UV} laser for improved and novel
  experiments on hydrogen,} Ph.D. thesis, Colorado State University (2019).

\bibitem{253nmBBO}
R.~Zhao, Z.~Fu, L.~Zhang, S.~Fang, J.~Sun, Y.~Feng, Z.~Xu, and Y.~Wang,
  \enquote{High-power continuous-wave narrow-linewidth 253.7 nm
  deep-ultraviolet laser,} {\protect\JournalTitle{Applied Optics}} \textbf{56},
  8973--8977 (2017).

\bibitem{229nmBBOCdMOT}
Y.~Kaneda, J.~M. Yarborough, Y.~Merzlyak, A.~Yamaguchi, K.~Harashida, N.~Ohmae,
  and H.~Katori, \enquote{Continuous-wave, single-frequency 229 nm laser source
  for laser cooling of cadmium atoms,} {\protect\JournalTitle{Opt. Lett.}}
  \textbf{45}, 705--708 (2016).

\bibitem{CLBOphotorefractivedamageAl}
K.~Takachiho, M.~Yoshimura, Y.~Fukushima, Y.~Takahashi, M.~Imade, T.~Sasaki,
  and Y.~Mori, \enquote{Al doping of {C}s{L}i{B}$_6${O}$_{10}$ for high
  resistance to ultraviolet-induced degradation,}
  {\protect\JournalTitle{Applied Physics Express}} \textbf{6}, 022701 (2013).

\bibitem{CLBOaldoping2017}
X.~Zhu, H.~Tu, Y.~Zhao, and Z.~Hu, \enquote{Growth of high quality {A}l-doped
  {C}s{L}i{B}$_6${O}$_{10}$ crystals using {C}s$_2${O}-{L}i$_2${O}-{M}o{O}$_3$
  fluxes,}  (2017).

\bibitem{Continuousbeam}
J.~C. Shaw and D.~J. McCarron, \enquote{Bright, continuous beams of cold free
  radicals,} {\protect\JournalTitle{Phys. Rev. A}} \textbf{102}, 041302 (2020).

\bibitem{HallHeNe}
T.~M. Niebauer, J.~E. Faller, H.~M. Godwin, J.~L. Hall, and R.~L. Barger,
  \enquote{Frequency stability measurements on polarization-stabilized
  {H}e-{N}e lasers,} {\protect\JournalTitle{Appl. Opt.}} \textbf{27},
  1285--1289 (1988).

\bibitem{AlClhfs}
J.~Hoeft, T.~Törring, and E.~Tiemann, \enquote{Hyperfeinstruktur von {A}l{C}l
  and {A}l{B}r / hyperfine structure of {A}l{C}l and {A}l{B}r,}
  {\protect\JournalTitle{Zeitschrift fur Naturforschung}} \textbf{28A},
  1066--1068 (1973).

\bibitem{CLBOdegradation2}
M.~Yoshimura, Y.~Takahashi, H.~Adachi, and Y.~Mori, \enquote{Nonlinear crystals
  for deep-{UV} light generation,} in \emph{Advanced Solid State Lasers,}
  (Optical Society of America, 2014), p. ATu4A.1.

\bibitem{15mmCLBO}
J.~Sakuma, Y.~Asakawa, T.~Imahoko, H.~Sekita, and M.~Obara, \enquote{High-power
  continuous-wave 266 and 213~nm generation with {C}s{L}i{B}$_6${O}$_{10}$
  crystals,} in \emph{Conference on {L}asers and
  {E}lectro-{O}ptics/{I}nternational {Q}uantum {E}lectronics {C}onference and
  {P}hotonic {A}pplications {S}ystems {T}echnologies,}  (Optical Society of
  America, 2004), p. CTuI1.

\end{thebibliography}

%%%%%%%%%% If preparing manually:
% \newcommand{\enquote}[1]{``#1''}

% \bibitem{Zhang:14}
% Y.~Zhang, S.~Qiao, L.~Sun, Q.~W. Shi, W.~Huang, L.~Li, and Z.~Yang,
%   \enquote{Photoinduced active terahertz metamaterials with nanostructured
%   vanadium dioxide film deposited by sol-gel method,}
%   {\protect\JournalTitle{Optics Express}} \textbf{22}, 11070--11078 (2014).

% {Optical Society}, \enquote{{OSA Publishing},}
%   \url{http://www.osapublishing.org}.

% \bibitem{FORSTER2007}
% P.~Forster, V.~Ramaswamy, P.~Artaxo, T.~Bernsten, R.~Betts, D.~Fahey,
%   J.~Haywood, J.~Lean, D.~Lowe, G.~Myhre, J.~Nganga, R.~Prinn, G.~Raga,
%   M.~Schulz, and R.~V. Dorland, \enquote{Changes in atmospheric consituents and
%   in radiative forcing,} in \enquote{Climate Change 2007: The Physical Science
%   Basis. Contribution of Working Group 1 to the Fourth assesment report of
%   Intergovernmental Panel on Climate Change,}  S.~Solomon, D.~Qin, M.~Manning,
%   Z.~Chen, M.~Marquis, K.~B. Averyt, M.~Tignor, and H.~L. Miler, eds.
%   (Cambridge University Press, 2007).

% \end{thebibliography}

\end{document}